\documentclass{article}%
\usepackage{amssymb}
\usepackage{amsmath}
\usepackage{amsfonts}
\usepackage{graphicx}%
\begin{document}
{\LARGE Thomas-Fermi Screening in Graphene}%

\[
\]
M. Salis$^{a}$

\textit{Dipartimento di Fisica - Universit\`{a} di Cagliari- Cittadella
Universitaria , 09042 Monserrato-Cagliari, Italy}%

\[
\]
\textbf{Abstract} - The in-plane static screening of the field originated by a
charge placed in a graphene sheet is investigated. A self-consistent field
equation in the real space domain is obtained by using a suitable Thomas-Fermi
procedure. Exact and approximated (for qualitative considerations) solutions
are presented. In the case of a charged sheet, the screened potential presents
a tail dependent on the free carrier density whose importance is connected
with the local features of the impurity system. Early conclusions about
Thomas-Fermi screening in graphene are revised.

PACS: 71.10.Ca, 73.61.Le%
\[
\]

Graphene is a bidimensional structure which has newly attracted the interest
of physic community since the recent report on the successful isolation of
free-standing carbon monolayers [$1$]. The striking properties of this
material ensue from the peculiar electronic structure which in the ideal
honeycomb lattice shows gapless (approximately) linear-spectrum bands with
degeneracy points (two per cell) at the corners of the hexagonal Brillouin
zone [$2$]. Due to a peculiar suppression of localization effects, electrical
conductivity never falls below a certain value (resistivity $\rho_{\max
}\approx6.5~k\Omega$) corresponding to the quantum unit of Mott conductance
[$1$]. Anomalous effects in transport properties were successfully explained
in the framework of the quantum electrodynamics [$3$] which idealize the
electron spectrum (near the two degeneracy points) as the one of a 2D-gas of
massless two-fermion-Dirac quasiparticles [$4$]. Carrier populations can be
changed by thermal excitation or/and by electrically induced band shift.
Electron or holes are poured into bands in order to match the equilibrium
Fermi level. Accordingly, electron or hole conductivities can be settled by
applying suitable gate voltages [$1$].

This paper is devoted to the investigation of the static screening properties
in the gapless bidimensional semiconductor. \ Really, screening in
bidimensional electron gas was already investigated by using the
self-consistent-field dielectric formulation of Ehrenreich and Cohen (EC)
[$5$]. According to the EC's work, involved in 3D electron gas, the self
consistency can be attained by combining the single-particle Liouville and the
Poisson equations [$6$], thus allowing for the Lindhard \ longitudinal
dielectric constant [$7$]. The repetition of this procedure in a 2D system
[$5$] may be questioned since it lacks a proper handling of Poisson equation
for the confined electron gas. A different approach, still based on the EC's
work, was addressed to investigate static screening in graphite layers [$8$].
To attain the self-consistency, the static solution of the Liouville equation
was inserted in the calculus of the induced potential. Layers was treated as
bidimensional electron systems each having a disk-shaped Fermi
(parabolic-spectrum) distribution with radius $k_{F}$. Both the Thomas-Fermi
(TF) screening and the Friedel oscillations were calculated. The authors
concluded that, due to the electron confinement, the TF screening in the layer
is independent of $k_{F}$ and that it is asymptotically ruled by a law such as
$V(x)\sim A~\exp(-Bx)/x$ where $A$ and $B$ are suitable constants. The case of
vanishingly small carrier density was handled in the short-range limit by
assuming that the (RPA) screening come out from all the $\pi$ electron (one
electron per atom). Besides the latter model assumption, we believe that
conclusions about features of the TF screening , that is the independence of
the carrier density are to be reconsidered.

Differently from early works, the TF screening in the graphene sheet will be
investigated in the real space domain. To properly account for the background
polarization the screening problem will be addressed on the line of the
Oliva's model [$9$]. This model extends the TF theory to include
semiconductors or insulators in which a perturbing field succeeds in
"piercing" the electron forbidden gap, thus allowing valence-band electrons to
penetrate the conduction band. Accordingly, the local excess of electron
density is determined by the band-shift across the higher occupied energy
level. To avoid unnecessary complications we disregard effects due to
polarization of the medium surrounding. Of course, the use of the TF approach
leaves out effects such as Friedel oscillations which become overwhelming at
large distance from the source of the perturbing field. We will consider both
the cases of a charged (by gate voltage) and uncharged graphene sheet. The
screening problem leads to a non-linear integro-differential equation, but
essential features of the screened potential can be deduced from a simplified
form of such an equation. It will be shown that the decay of the potential
depends on the carrier density in the graphene sheet.

Let us consider a large (virtually infinite) carbon monolayer at the
temperature $T=0~K$ in which a gate voltage produces an electron-level energy
shift $-eU<0$ ($-e<0$). The layer is thus charged by an uniform electron
distribution with density $n^{\ast}$ . If a point-like positive charge
particle if placed in such a bidimensional system, the electron distribution
adjust itself to reach the minimum energy state with a density $n(r)$
dependent on the distance from the external charge. The induced change of
charge density is thus $-e\delta n(r)=-e\left(  n(r)-n^{\ast}\right)  $. For
simplicity's sake \ we will consider the bidimensional system as a continuum
\ with circular isotropic properties. The dipole background polarization
effect is phenomenologically subsumed in a constant $\epsilon$. We will return
later on the afore presented model assumptions. Now, if we fix the\ the origin
of the polar coordinate system on the external charge, the induced potential
can be calculated as%
\[
V_{ind}\left(  r\right)  =-\frac{e}{\epsilon}%
{\displaystyle\int\limits_{0}^{2\pi}}
{\displaystyle\int\limits_{0}^{\infty}}
\frac{R}{\sqrt{R^{2}-2rR\cos\vartheta+r^{2}}}\delta n(R)~~d\vartheta dR~.
\]
For $r>0$ we have%

\begin{equation}
V_{ind}\left(  r\right)  =-\frac{e}{\epsilon}r%
{\displaystyle\int\limits_{0}^{\infty}}
p(x)\delta n(xr)~~dx~ \label{TF1}%
\end{equation}
where%

\begin{equation}
p(x)=\int x(x^{2}-2x\cos\vartheta+1)^{-1/2}d\vartheta\label{TFXX1}%
\end{equation}
which shows $p(0)=0$ ($dp(x=0)/dx=2\pi$), $p(\infty)=2\pi$ ($dp(x=\infty
)/dx=0$) and a singularity at x=1. Since $\delta n(r)$ is absolutely
integrable, the integral function (\ref{TF1}) is continuous and its
derivatives can be obtained by differentiating under the integral sign. To
avoid ambiguities in handling singular functions, we replace the function
$p(x)$ with a function $\overline{p}(x)$ which differ from the former only in
a suitably small interval around x=1 where it is finite and continuous as well
as its derivatives. It is not necessary to specify the exact form of
$\overline{p}(x)$ , we only require that $-(1/2\pi)\int_{0}^{\infty}\left(
d^{2}\overline{p}(x)/dx^{2}\right)  dx=1$ compatibly with the total
charge-screening condition, that is, $2\pi\int_{0}^{\infty}\delta
n(r)~r\ dr=1$. On the other hand, we should remove the singularity to avoid
considering divergent self-interactions. By defining $V\left(  r\right)
=e/r+V_{ind}\left(  r\right)  $ it is easy to verify that ($r>0$).%

\begin{equation}
D_{r}V\left(  r\right)  =\frac{1}{r}\frac{d}{dr}r^{2}\frac{d}{dr}V\left(
r\right)  =-\frac{e}{\epsilon}\left(  2+4r\frac{d}{dr}+r^{2}\frac{d^{2}%
}{dr^{2}}\right)
{\displaystyle\int\limits_{0}^{\infty}}
p(x)\delta n(xr)~~dx~. \label{TF2}%
\end{equation}
We will simplify this equation later. Now, we search for the self-consistency
by relating $\delta n(r)$ to $V(r)$ , that is [$9$],%

\begin{equation}
\delta n(r)=2\gamma_{g}%
{\displaystyle\int\limits_{-\infty}^{\infty}}
\frac{\left\vert \varepsilon\right\vert }{\exp\left[  (\varepsilon-eV\left(
r\right)  -eU)/kT\right]  +1}d\varepsilon
\;\ \ \ \ \ \ \ \ \ \ \ \ \ \ \ \ \ \ \ \ \ \ \ \ \ \ \ \ \ \ \ \ \ \ \ \ \ \label{TF3BIS}%
\end{equation}%
\[
\;\ \ \ \ \ \ \ \ \ \ \ \ \ \ \ \ \ \ \ \ \ \ \ \ \ \ \ \ \ \ \ \ \ \ \ \ \ \ \ \ \ \ -2\gamma
_{g}%
{\displaystyle\int\limits_{-\infty}^{\infty}}
\frac{\left\vert \varepsilon\right\vert }{\exp\left[  (\varepsilon
-eU)/kT\right]  +1}d\varepsilon
\]
where we used the density of states $N(\varepsilon)=4\left\vert \varepsilon
\right\vert /3\pi\gamma_{0}^{2}a^{2}$ ($\gamma_{0}\approx1$ $eV$, $a=2.46A$)
[$3$] and $\gamma_{g}=4/3~\pi~\gamma_{0}^{2}a^{2}\approx0.07~eV^{-2}A^{-2}$
The infinite integration range is allowed by the small energies involved in
the problem dealt with. Really, we are also assuming that the electronic
energy spectrum maintains its linear law over a large range of energies
[$1,3$]. At $T=0$ we obtain%

\begin{equation}
\delta n(r)=e^{2}\gamma_{g}\left\{  \left[  V\left(  r\right)  +U\right]
^{2}-U^{2}\right\}  \label{TF4}%
\end{equation}
In the following, we will present the screened potential in the more
convenient form
\begin{equation}
V(r)=e~\frac{f(r)}{r} \label{TF12}%
\end{equation}
where the screening factor satisfies the conditions $f(0)=1/\epsilon$ and
$f(\infty)=0$. Thus, after substitution of eq. (\ref{TF4}) into eq.
(\ref{TF2}) and integration by parts we obtain%

\begin{equation}
\frac{d^{2}}{dr^{2}}f(r)=\frac{2\pi e^{4}\gamma_{g}}{\epsilon r}%
{\displaystyle\int\limits_{0}^{\infty}}
\varphi(x)\left(  \frac{f^{2}\left(  \xi\right)  }{\xi}+2\frac{U}{e}f\left(
\xi\right)  \right)  _{\xi=xr}dx\label{TF13}%
\end{equation}
where the function $\varphi(x)=-(1/2\pi)xd^{2}\overline{p}(x)/dx^{2}$ is
everywhere small ($\varphi(0)=0,\varphi(\infty)=0$) and negative besides a
close interval around $x=1$ where it oscillates and shows its maximum positive
value at $x=1$. Due to these properties, in some favorable cases (very small
$U$)  the solution of eq. (\ref{TF13}) does not greatly deviates from the one
obtained by replacing $\varphi(x)$ with \ the Dirac function $\delta(x-1)$.
This allows us to investigate qualitatively some properties of the screening
factor which will be otherwise hidden by the numerical calculus. Thus, only
for qualitative discussion, we consider%

\begin{equation}
\frac{d^{2}}{dr^{2}}f(r)=\frac{2\pi e^{4}\gamma_{g}}{\epsilon}\left(
\frac{f^{2}\left(  r\right)  }{r^{2}}+2\frac{U}{e}\frac{f\left(  r\right)
}{r}\right)  ~. \label{TF14}%
\end{equation}
Equation (\ref{TF14}) can be analytically solved in the linear range, that is,
in the range of distances where $\left\vert V\left(  r\right)  \right\vert
<<\left\vert U\right\vert $ holds so that we can write%

\begin{equation}
\frac{d^{2}}{dr^{2}}f(r)=\frac{4\pi e^{3}\gamma_{g}U}{\epsilon}\frac{f\left(
r\right)  }{r}~. \label{TF15}%
\end{equation}
The solution of the linearized equation is [$10$]%
\begin{equation}
f\left(  r\right)  =\Lambda\sqrt{\left(  4\lambda r\right)  }K_{1}\left(
\sqrt{4\lambda r}\right)  \label{TF6}%
\end{equation}
where $K_{1}\left(  x\right)  $ stands for the modified Bessel (Basset)
function of the second kind, $\lambda=4\pi e^{3}\gamma_{g}U/\epsilon=\left(
eU/\gamma_{0}\right)  31/a\epsilon$ is the inverse of the "screening length"
and $\Lambda$ is a factor merging the short and long range solutions. We
should take care in defining the screening length since it may happen that
when linearization holds the potential is significantly damped by the short
range screening. Really, when linearization holds we can use the asymptotic
form of the Bessel function, that is [$11$],%

\begin{equation}
f\left(  r\right)  =\Lambda\left(  4\lambda r\right)  ^{1/4}\exp
(-\sqrt{4\lambda r})~.\label{TF7}%
\end{equation}

In the short range distances where $\left\vert V\left(  r\right)  \right\vert
>>\left\vert U\right\vert $ the form of $f(r)$ is independent of $U$, that is,
independent of free carrier density. For language convenience, we will refer
to this case as the intrinsic screening. In the case $U=0$ , eq.. (\ref{TF14}) becomes%

\begin{equation}
\frac{d^{2}f(r)}{dr^{2}}=\alpha\left[  \frac{f(r)}{r}\right]  ^{2} \label{TF9}%
\end{equation}
where $\alpha=2\pi e^{4}\gamma_{g}/\epsilon\simeq91/\epsilon$. We are able to
give an asymptotic solution of this equation. Indeed, for $f(r)<<1$ we can write%

\begin{equation}
\frac{df(r)}{dr}=-\frac{\alpha}{r}\left\{  f(r)^{2}-2\alpha f(r)^{3}+O\left[
f(r)^{4}\right]  \right\}  \label{TF10}%
\end{equation}
which, when higher order term can be disregarded, has the solution%

\begin{equation}
f(r)=1/\alpha\ln(r/b) \label{TF11}%
\end{equation}
where $b$ is a suitable length constant whose value depend on the dielectric
constant. From eq. (\ref{TF11}) it is clear\ that screening becomes more
efficient as $\epsilon$ decreases. If we calculate the screening factor for
two different dielectric constant $\epsilon_{1}$ and $\epsilon_{2}$ with
$\epsilon_{2}>\epsilon_{1}$, at (very large) distances $r$ where the
asymptotic form holds, we have $f_{1}(r)/f_{2}(r)\approx\epsilon_{1}%
/\epsilon_{2}$.

Equation (\ref{TF9}) is clearly scale-factor free since $\alpha$\ is
dimensionless. This special property cannot be ascribed only to the
bidimensional character of the problem dealt with. But, it also come out from
the linearity of the electronic spectrum. As a mere speculation, an analogous
equation would play in the case of electroweak screening by the neutrino sea
or, in general, by 3D gapless linear-spectrum fermion systems (the screening
factor will appear as $f(r)^{3}$). \ Unfortunately, this property makes the
screening factor sensitive of the impurity-system features. For more clarity,
let us assume that the external charge has a radius $u$ . We change the
boundary conditions as $f(u)=1/\epsilon$. Thus, the potential in the intrinsic
screening case can be written also as%

\begin{equation}
V(r)=e~\frac{f(x)}{ux} \label{TF16}%
\end{equation}
where $x=r/u\geq1$. \ From comparison between the potential (\ref{TF16}) and
and the level shift induced in the charged graphene sheet we can estimate the
reduced distance $x^{\ast}$ where the carrier density effect becomes
significant, that is, $x^{\ast}\sim ef(x^{\ast})/Uu$ . Thus, if $u\rightarrow
0$ we have $x^{\ast}\rightarrow\infty$ and $f(x^{\ast})\rightarrow0$
(vanishingly small $\Lambda$ values). In few words, the importance of free
carrier density decreases as $u$ decreases. In this connection, it is to be
pointed out that several model assumptions fails as $u\rightarrow0$, mainly:
1) the screening is not bidimensional near the external charge; 2) the
electronic spectrum is not linear over the energies spanned by the large
$V(r)$; 3) the system cannot be considered as a continuum. However, these are
general troubles and we can reasonably assume that the impurity as a radius
not too small, namely $u=1\mathring{A}$. \ The results are not qualitatively
affected if we use the half or the twice of this value.

For the numerical integration of eq. (\ref{TF13}) we used the constraints on
$\overline{p}(x)$ or, equivalently, the condition $\int_{0}^{\infty}%
\varphi(x)dx=0$. Figure 1 shows the results for the cases: $\epsilon=10$,
$U=5~10^{-4}~V$ (solid curve A), $U=5~10^{-6}$ $V$ \ (solid curve B) and $U=0$
\ (solid curve B). We used small $U$ values to facilitate numerical calculus.
The dashed curves represent the corresponding exact solutions of eq.
(\ref{TF14}). The asymptotic expressions (\ref{TF7}) and (\ref{TF11})
($\Lambda=0.91$ and $\Lambda=0.033$ for A and B , respectively, and
$b=7.54~\mathring{A}$) are shown by means of dotted curves. It appears that
deviations between solutions (\ref{TF13}) and (\ref{TF14}) increase as $U$
increases $\left[  13\right]  $.

As a final note, we briefly consider the temperature effect on the screening
charge density the case $\left\vert eV\left(  r\right)  \right\vert
<<\left\vert eU\right\vert <<kT$. Now, we must take into account that the
induced charge density $-e\delta n(r)$ includes the two contributions
$-e\delta n_{-}(r)$ \ and $-e\delta n_{+}(r)$ arising from electrons and
holes, respectively. It is easy to verify that eq. (\ref{TF3BIS}) satisfies
$\delta n(r)=\left[  \delta n_{-}(r)-\delta n_{+}(r)\right]  $\ . Thus, by
expanding the integral with respect to $(eV(r)+eU)/kT$ \ and by substituting
in eq. (\ref{TF2}) , we obtain%

\begin{equation}
\delta n(r)\approx2e\gamma_{g}\left(  \xi kT+\kappa eU\right)  V\left(
r\right)  \label{T12}%
\end{equation}
where, by putting $g(x)=\left\vert x\right\vert \exp(x)/\left(  \exp
(x)+1\right)  ^{2}$, $\xi=\int_{-\infty}^{+\infty}g(x)~dx$ and $\kappa
=\int_{-\infty}^{+\infty}g(x)~\left(  2\exp(x)-1\right)  /\left(
\exp(x)+1\right)  dx$. Note that In the case $U=0$ the screening parameter
increases proportionally to the temperature, that is, $\lambda(T)=4\pi
e^{2}\gamma_{g}\xi kT/\epsilon$. The exact numerical calculus of the screening
factor appear at this level an unnecessary complication. Thus, we do not dwell
upon this point..%
\[
\]
\textbf{References}

$^{a}$masalis@unica.it

[$1$] K. S. Novoselov, A. K. Geim, S. V. Morozov, D. Jiang, M. I. Katshelson,
I. V. Grigorieva, S. V. Dubonos and A. A. Firsov, Nature \textbf{438}, (2005)
197; K. S. Novoselov, A. K. Geim, S. V. Morozov, D. Jiang, Y. Zhnag, I. V.
Grigorieva, S. V. Dubonos and A. A. Firsov, Science \textbf{306}, (2004) 666.

[$2$] P. R. Wallace, Phys. Rev. B \textbf{71} (1947) 622.

[$3$] N. M. Peres, F. Guinea, A. H. Castro Neto, Phys. Rev B \textbf{73},
125411 (2006); V. P. Gusynin and S. G. Sharapov, Phys. Rev. B \textbf{73}
(2006) 245411; 

[$4$] G. W. Semenoff, Phys. Rev. Lett \textbf{24} (1984) 2449.

[$5$] F. Stern, Phys. Rev. Lett. , \textbf{18} (1967) 546.

[$6$] \ H Eherenreich and M. H. Cohen, Phys. Rev. \textbf{115} (1959) 786.

[$7$] G. D. Mahan, Many Particle Physics (Plenum Press, New York-London) 1981,
Ch. 5

[$8$] P. B. Visscher and L. M. Falicov, Phys. Rev. B \textbf{3} (1971) 2541.

[$9$] J. Oliva, Phys. Rev. \textbf{B} \textbf{35} (1987) 3431.

[$10$] I. S. Gradshteyn and I. M. Ryzhik, Tables of integrals, Series and
Products, Sixth Edition \ (Academic Press) 2000, p. 922 n.5

[$11$] E. T. Wittaker and G. N. Watson, A Course of Modern Analysis (Cambridge
at the University Press) 1952, p. 374

$\left[  13\right]  $ When writing this article the author was not aware of a
similar paper published in november 2006: M. I. Katnelson, Phys. Rev. B
\textbf{74}, 201401 (2006). By applying the Fourier transforms to a linearized
integral self-consistent-field equation it was found that $f(r)$ decreases
asymptotically as $\sim1/r^{2}$ which, substantially, confirms the result of
ref. $\left[  5\right]  $.%

\[
\]

\[
\]

\[
\]

\[%
{\includegraphics[
natheight=8.500200in,
natwidth=11.000400in,
height=3.9704in,
width=5.1301in
]%
{GraficiOrigin/Fig1Art.bmp}%
}%
\]
\textbf{Caption of Fig. 1}

Screening factor dependence on the distance from the external charge with
radius $u=1\mathring{A}$ in a sheet with $\epsilon=10$. Solid curves represent
the exact numerical solutions of (\ref{TF13}): $U=5~10^{-4}~V$ (\textbf{A)},
$U=5~10^{-6}~V$ (\textbf{B) }and $U=0$ (\textbf{C)}. Dashed curves represent
the exact numerical solutions of eq. (\ref{TF14}): $U=5~10^{-4}~V$
(\textbf{A)}, $eU=5~10^{-6}~V$ (\textbf{B) }and $U=0$ (\textbf{C)}. Dotted
curves represent the asymptotic solution of eq. (\ref{TF14}) with
$\Lambda=0.91$ (\textbf{A), }$\Lambda=0.033$ (\textbf{B) }and $b=7.54$
$\mathring{A}$. (\textbf{C). }

\end{document}